\documentclass[showpacs,pre]{revtex4}
\usepackage{dcolumn}
\usepackage{amssymb}
\usepackage{amsmath}
\usepackage{graphicx}
\usepackage{subfigure}
\usepackage{caption2}

\begin{document}

\title{Partially incoherent gap solitons in Bose-Einstein condensates}
\author{I. M. Merhasin$^{1}$, Boris A. Malomed$^{2}$ and Y. B. Band$^{3}$}
\address{$^1$Department of Electrical and Electronics Engineering, The College of Judea and Samaria,
Ariel,
Israel\\
$^2$Department of Interdisciplinary Studies, School of Electrical
Engineering, Faculty of Engineering, Tel Aviv University, Tel Aviv
69978,
Israel\\
$^3$Departments of Chemistry and Electro-Optics, and The Ilse Katz
Center for Nano-Science, Ben-Gurion University, Beer Sheva 84105,
Israel }

\begin{abstract}
We construct families of incoherent matter-wave solitons in a repulsive
degenerate Bose gas trapped in an optical lattice (OL), i.e., gap solitons,
and investigate their stability at zero and finite temperature, using the
Hartree-Fock-Bogoliubov equations. The gap solitons are composed of a
coherent condensate, and normal and anomalous densities of incoherent vapor
co-trapped with the condensate. Both \textit{intragap} and \textit{intergap}
solitons are constructed, with chemical potentials of the components falling
in one or different bandgaps in the OL-induced spectrum. Solitons change
gradually with temperature. Families of intragap solitons are completely
stable (both in direct simulations, and in terms of eigenvalues of
perturbation modes), while the intergap family may have a very small
unstable eigenvalue (nevertheless, they feature no instability in direct
simulations). Stable higher-order (multi-humped) solitons, and bound
complexes of fundamental solitons are found too.
\end{abstract}

\maketitle

\section{Introduction\textit{\ }}

An ultracold Bose gas is, in general, a mixture of a coherent Bose-Einstein
condensate (BEC) and an incoherent (fluctuating) \textquotedblleft vapor".
As shown in Refs.~\cite{HFB,Vardi}, the Hartree-Fock-Bogoliubov (HFB)
description of such a mixture in one dimension is provided by the
Gross-Pitaevskii (GP) equation for the condensate order parameter $\phi (x,t)
$, coupled to equations for components of the vapor wave function, $u(x,t)$
and $v(x,t)$, which are responsible for normal and anomalous densities of
the fluctuations. This approach and related formalisms for the description
of fluctuations make it possible to analyze various effects, such as quantum
phase diffusion in BEC\ and its depletion in a time-dependent trap through
transfer of atoms to non-condensed states \cite{depletion}, depletion of
dark solitons \cite{depletion-dark}, quantum-noise squeezing of \textit{gap
solitons} in a repulsive condensate trapped in an optical lattice (OL) \cite%
{Canberra}, deviations from one-dimensionality \cite{Luca}, and friction and
diffusion of solitons in a cloud of thermal atoms \cite{friction}. The full
system of coupled HFB equations was used in Ref.~\cite{Vardi} to show that,
in the case of attraction between atoms, the matter flux from the condensate
to the vapor may lead to splitting of bright matter-wave solitons (which
have been created experimentally in $^{7}$Li \cite{exper-bright} and $^{85}$%
Rb \cite{Cornish} condensates) into two fragments that may be regarded as
\textit{partially incoherent} solitons, i.e., bound states of the coherent
condensate and vapor components, similar to partially incoherent solitons
known in nonlinear optics \cite{Moti}.

The objective of this work is to find solutions for partially incoherent gap
solitons (GSs) in an OL potential, and investigate their stability, at zero
and finite temperatures. GSs in BEC were predicted using the GP equation
\cite{GSprediction}, and then created experimentally in a $^{87}$Rb
condensate \cite{gap_sol} (these solitons were quite \textquotedblleft
meager", each consisting of just a few hundred atoms). Partially incoherent
lattice solitons at finite temperature ($T$) were qualitatively predicted in
Ref. \cite{Ahufinger}, which relied upon simulations of the GP equation,
starting with a random Bose distribution at finite $T$ and gradually forming
a soliton by switching the OL potential on. Families of two-component GSs of
\textit{intragap} and \textit{intergap} types (see below), i.e., with
chemical potentials of the components falling in one or different bandgaps
of their common OL-induced linear spectrum, were found~within the framework
of GP equations coupled by repulsion between the species \cite{Gubeskys}. It
is relevant to mention that two-component solitons of the intergap type \cite%
{Moti2}, as well as their discrete counterparts \cite{Sukhorukov}, were
earlier predicted in lattice models of nonlinear optics \cite{Moti};
however, these were coherent objects and, unlike the model considered here
(see below), they were found in models that did not include coherent
four-wave-mixing terms.

The GS is not a ground state of the repulsive condensate (it is obvious that
it does not realize an absolute energy minimum for the self-repulsive
condensate loaded into the OL potential, with a given number of atoms), but
it is nevertheless stable. It is relevant to stress the difference of the
analysis of quantum fluctuations around the GSs from the problem of quantum
depletion of dark solitons, where fluctuations fill out the notch at the
center of the soliton and thus gradually destroy it \cite{depletion-dark}.
For GSs (which are bright solitons), the notch is absent in the family of
solutions found in the first bandgap [see Fig.~\ref{Fig1}(a) below]. In the
second gap, notch(es) may be present in decaying wings of the soliton's
waveform [see Figs.~\ref{Fig1}(b) and \ref{Fig4}], but the soliton's
identity is not predicated on them.

The paper is organized as follows. In Section II, we formulate the model,
which is based on time-dependent HFB equations for the coherent condensate
and incoherent vapor components interacting with it (at finite $T$). Basic
results for the partially coherent GS families of the intra- and intergap
types, found in a numerical form at $T=0$ and $T>0$, are reported in Section
III. Examples of higher-order (multi-humped) solitons of various types are
presented in Section III too. Stability of these solitons is investigated,
by means of computation of the corresponding eigenvalues for small
perturbations, and in direct simulations, in Section IV. Section V concludes
the paper.

\section{The model}

Coupled time-dependent HFB equations are obtained as a truncation of a
hierarchy of approximations \cite{HFB} developed for the description of the
dynamics of interacting condensate and vapor components of the degenerate
bosonic gas, at very low but, generally, finite $T$. Following Ref.~\cite%
{Vardi}, the equations for the gas with repulsion between bosonic atoms are
cast in the following normalized form, with the nonlinearity coefficient
scaled to unity:
\begin{gather}
i\frac{\partial \phi }{\partial t}=-\frac{1}{2}\frac{\partial ^{2}\phi }{%
\partial x^{2}}+\left[ -\varepsilon \cos (2x)+\left\vert \phi \right\vert
^{2}+2\tilde{n}\right] \phi +\tilde{m}\phi ^{\ast },  \label{phi} \\
i\frac{\partial }{\partial t}\left(
\begin{array}{c}
u \\
v%
\end{array}%
\right) =\left(
\begin{array}{cc}
\hat{L} & -\tilde{m}-\phi ^{2} \\
\tilde{m}^{\ast }+(\phi ^{\ast })^{2} & -\hat{L}%
\end{array}%
\right) \left(
\begin{array}{c}
u \\
v%
\end{array}%
\right) .  \label{HFB_PDE}
\end{gather}%
Here $\hat{L}\equiv -(1/2)\partial _{xx}^{2}-\varepsilon \cos (2x)+2\left(
\left\vert \phi \right\vert ^{2}+\tilde{n}\right) $, $\tilde{n}\equiv \left(
1+\mathcal{N}_{B}\right) \left\vert v\right\vert ^{2}+\mathcal{N}_{B}|u|^{2}$
and $\tilde{m}\equiv -\left( 1+2\mathcal{N}_{B}\right) uv^{\ast }$ are the
normal and anomalous fluctuation densities (the asterisk stands for the
complex conjugation),
\begin{equation}
\mathcal{N}_{B}=\left( e^{E/kT}-1\right) ^{-1}  \label{Bose}
\end{equation}
is the Bose occupancy, with $E$ the ground-state energy of the tight
transverse confinement (note that the one-dimensional equations are derived
from their 3D counterparts, assuming strong confinement in the transverse
plane, by means of various approaches implying averaging in the transverse
plane \cite{3Dto1D,Lev}), and $\varepsilon $ is the strength of the
longitudinal OL potential, whose period is scaled to be $\pi $. A dynamical
invariant of Eqs.~(\ref{phi}-\ref{HFB_PDE}) is the total number of atoms,
\begin{equation}
N=\int_{-\infty }^{+\infty }\left[ 2\left\vert \phi (x)\right\vert
^{2}+\left( 1+2\mathcal{N}_{B}\right) \left( \left\vert u(x)\right\vert
^{2}+\left\vert v(x)\right\vert ^{2}\right) \right] dx\equiv N_{\phi
}(t)+N_{u}(t)+N_{v}(t)  \label{N}
\end{equation}%
(note that the expression for $N$ explicitly depends on $T$ via $\mathcal{N}%
_{B}$).

Partially incoherent GSs are looked for as bound states in which the
coherent condensate and incoherent vapor components are trapped together in
the OL,
\begin{equation}
\phi (x,t)=\Phi (x)e^{-i\mu _{\phi }t},\,\,\,\left\{ u(x,t),v(x,t)\right\}
=\left\{ U(x)\,e^{-i\mu _{u}t},V(x)\,e^{+i\tilde{\mu}_{v}t}\right\} ,
\label{HFB_subst}
\end{equation}%
with chemical potentials $\mu _{\phi }$, $\mu _{u}$ and $-\tilde{\mu}_{v}$
subject to the constraint (phase-locking condition),
\begin{equation}
\mu _{u}+\tilde{\mu}_{v}=2\mu _{\phi },  \label{locking}
\end{equation}
which implies that collisions may kick out pairs of condensate atoms into
the vapor (i.e., dynamical depletion of the condensate); the above-mentioned
transverse-confinement energy was subtracted from the chemical potentials.
Equations for the stationary parts of the wave functions are obtained by the
substitution of expressions (\ref{HFB_subst}) in Eqs. (\ref{phi}) and (\ref%
{HFB_PDE}):
\begin{eqnarray}
\mu _{\phi }\Phi  &=&-\Phi ^{\prime \prime }/2-\varepsilon \cos (2x)\Phi
+\left( \left\vert \Phi \right\vert ^{2}+2\tilde{n}\right) \Phi +\tilde{m}%
\Phi ^{\ast },  \notag \\
\mu _{u}U &=&-U^{\prime \prime }/2-\varepsilon \cos (2x)U+2\left( \left\vert
\Phi \right\vert ^{2}+\tilde{n}\right) U-(\Phi ^{2}+\tilde{m})V,  \notag \\
\tilde{\mu}_{v}V &=&-V^{\prime \prime }/2-\varepsilon \cos (2x)V+2\left(
\left\vert \Phi \right\vert ^{2}+\tilde{n}\right) V-\left( \Phi ^{2}+\tilde{m%
}\right) ^{\ast }U.  \label{HFB_ODE}
\end{eqnarray}

\begin{figure}[tbp]
\centering{\includegraphics[width=3.0in]{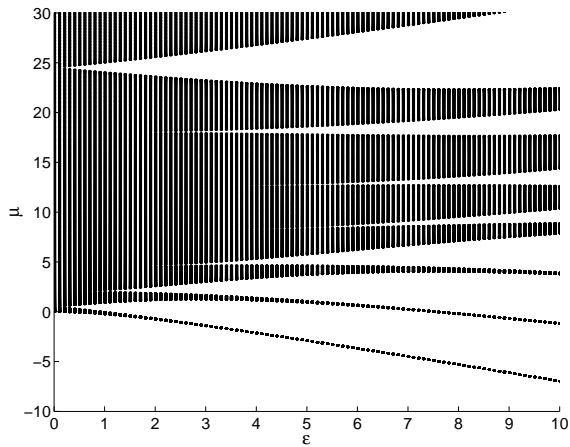}}
\caption{The bandgaps (unshaded areas) in the identical spectra of three
linearized equations (\protect\ref{HFB_ODE}), with $\protect\mu $ standing
for any chemical potential, $\protect\mu _{\protect\phi }$, $\protect\mu _{u}
$, or $\tilde{\protect\mu}_{v}$.}
\label{Fig0}
\end{figure}

In the linear approximation, Eqs.~(\ref{HFB_ODE}) decouple into the Mathieu
equation for the condensate, $\Phi ^{\prime \prime }+2\left[ \mu _{\phi
}+\varepsilon \cos (2x)\right] \Phi =0$, and its replicas for $U$ and $V$,
giving rise to identical bandgap spectra for the three waves. This bandgap
spectrum is displayed, for the sake of illustration, in Fig. \ref{Fig0}. GS
solutions can be found if $\mu _{\phi }$, $\mu _{u}$ and $\tilde{\mu}_{v}$
[the latter chemical potential is equal to $-\mu _{u}+2\mu _{\phi }$, as per
locking condition (\ref{locking})] belong to \emph{one} or \emph{different}
bandgaps. The respective solutions will be called intragap and intergap
solitons, as in Ref.~\cite{Gubeskys} (see also Refs. \cite{Moti2} and \cite%
{Sukhorukov}). In particular, in the typical case of a moderately strong OL,
which is used below to present generic results, with OL strength $%
\varepsilon =5$ (i.e., $2.5\,\,E_{R}$, where the recoil energy is, in
physical units, $E_{R}=\hbar ^{2}k^{2}/2m$, $k$ being the OL wavenumber),
the first and second bandgaps in Fig. \ref{Fig0} are $-2.5<\mu <0.7$ and $%
1.2<\mu <3.8$.

To conclude the description of the model, it is relevant to stress that the
fluctuations, $u(x)$ and $v(x)$, must include contributions from all Bloch
bands of the periodic potential. For sufficiently large $\varepsilon $, the
lowest bands are narrow, hence the Bloch states in each of them may be
approximated by a single mode in $u$ and $v$. For instance, inspection of
the spectrum from Fig. \ref{Fig0} for $\varepsilon =5$ (this value will be
used below) demonstrates that the three lowest bands are indeed sufficiently
narrow to be approximated by single modes, while other bands have little
relevance as they correspond to very high values of the chemical potential.
Furthermore, in this situation contributions from mode-mixing cross terms to
quadratic quantities (integrated densities, that measure the strength of the
fluctuation components in the Bose gas, see below) are negligible, in view
of the effective mutual incoherence of the Bloch wave functions in distinct
narrow bands separated by wide gaps. Therefore, in such a representation,
the integral quantities actually take the familiar form of diagonal sums
over several fluctuation modes \cite{HFB}-\cite{depletion-dark}.

\section{Numerical results: families of partially incoherent gap solitons}

\subsection{Intragap solitons}

\begin{figure}[tbp]
\centering\subfigure[]{\includegraphics[width=3.0in]{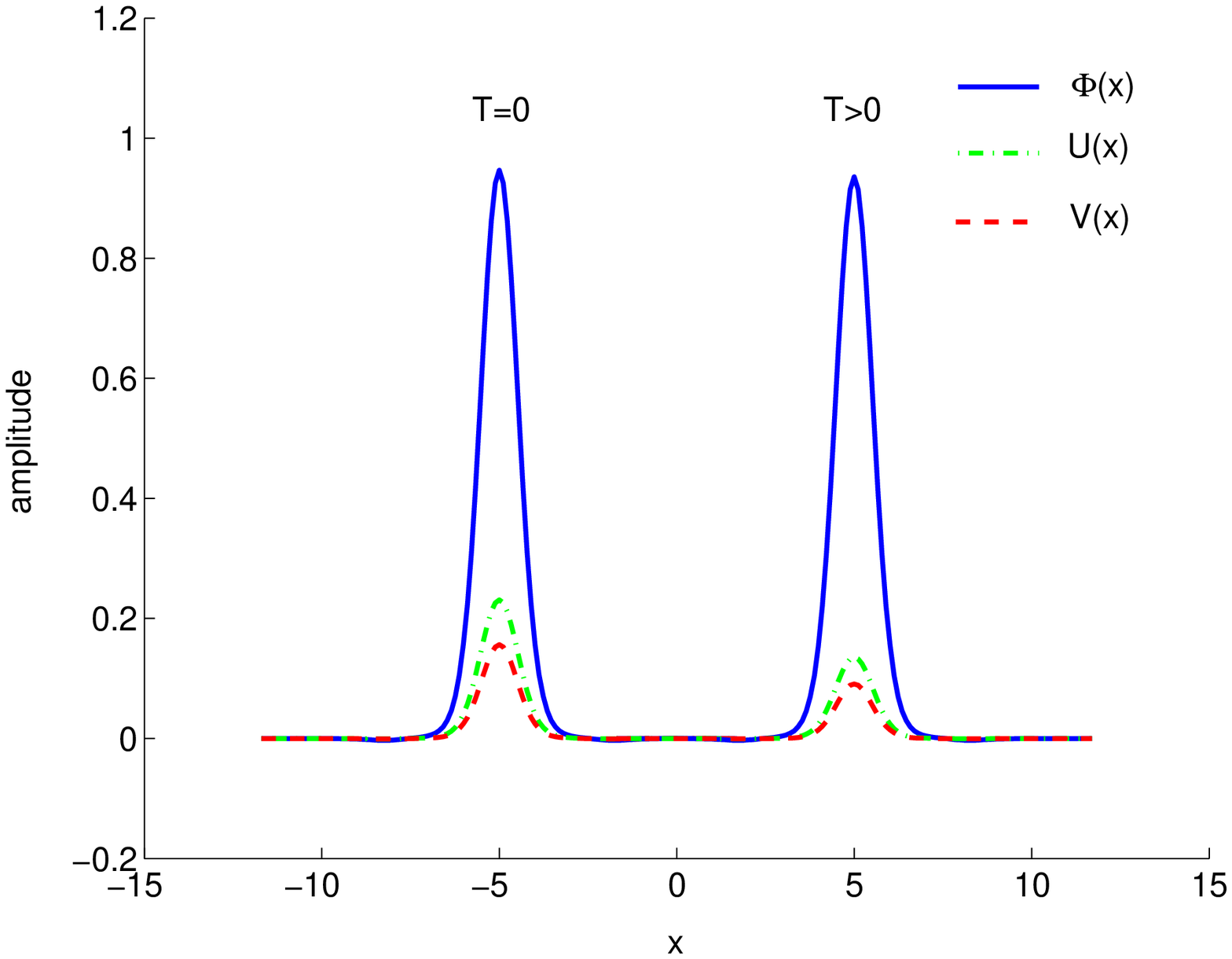}}
\centering\subfigure[]{\includegraphics[width=3.0in]{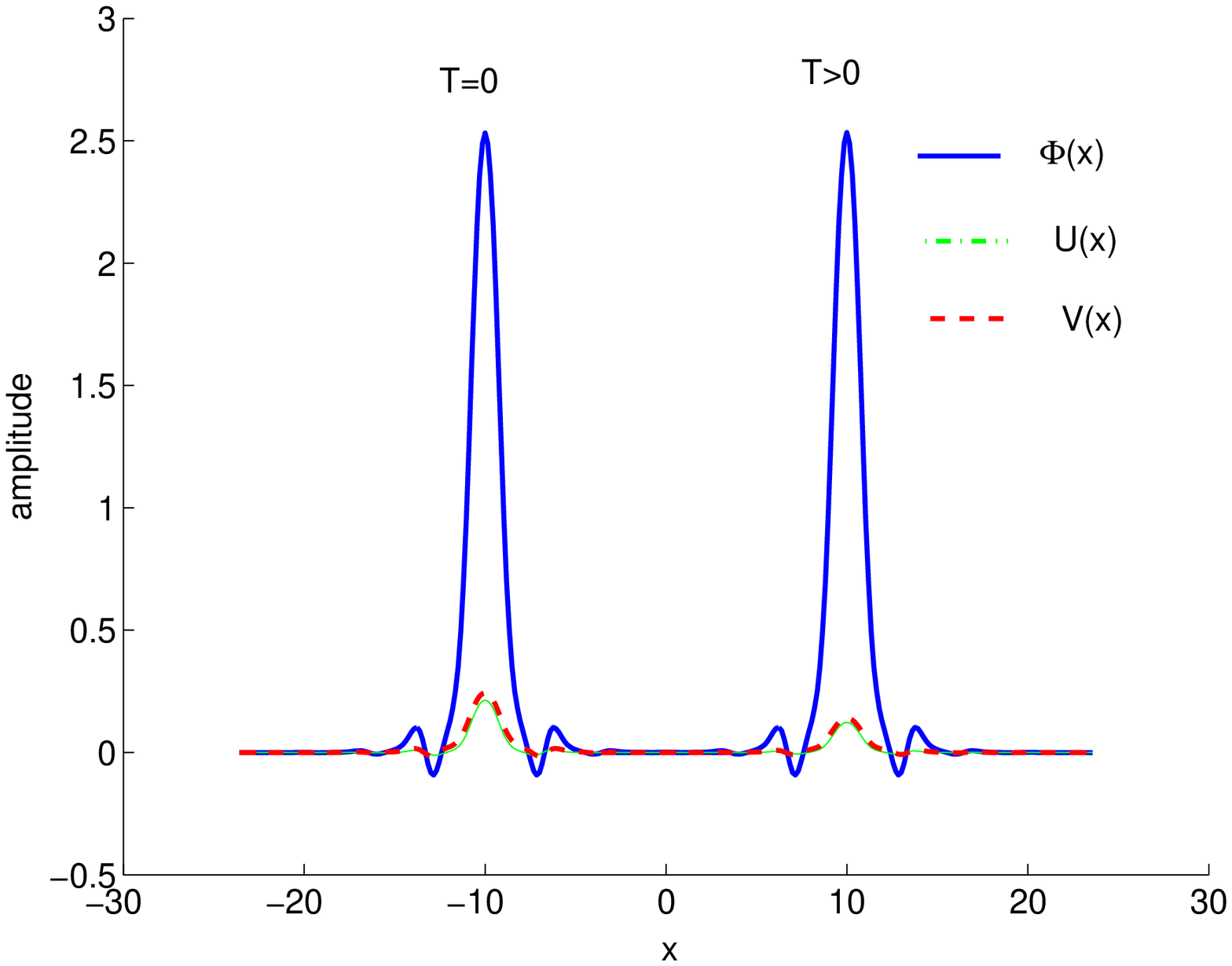}}
\caption{(Color online) Typical examples of stable intragap solitons found
for $T=0$ and $T>0$, which correspond to Bose occupancy $\mathcal{N}_{B}=0$
and $1$, respectively [see Eq. (\protect\ref{Bose})] in the first (a) and
second (b) bandgaps, with (a) $\protect\mu _{u}=-2$, $\tilde{\protect\mu}%
_{v}=-2.5$, $\protect\mu _{\protect\phi }=-2.25$; (b) $\protect\mu _{u}=1.5$%
, $\tilde{\protect\mu}_{v}=3$, $\protect\mu _{\protect\phi }=2.25$. In this
and other examples, the OL strength is $\protect\varepsilon =5$.}
\label{Fig1}
\end{figure}

Generic examples of solutions to Eqs.~(\ref{HFB_ODE}) with $T=0$ and $T>0$,
in the form of intragap GSs with the three chemical potentials falling in
the first or second bandgap, are shown in Fig.~\ref{Fig1}. In terms of Ref.~%
\cite{Gubeskys}, they are categorized as tightly and loosely bound solitons,
respectively. Note that $\mathcal{N}_{B}=1$ corresponds to $T\approx 70$ nK,
if the transverse trapping frequency is $2\pi \times 1~$KHz. In Fig.~\ref%
{Fig1}, it is seen that the soliton does not suffer drastic changes with the
increase of temperature.

\begin{figure}[tbp]
\centering\subfigure[]{\includegraphics[width=3.0in]{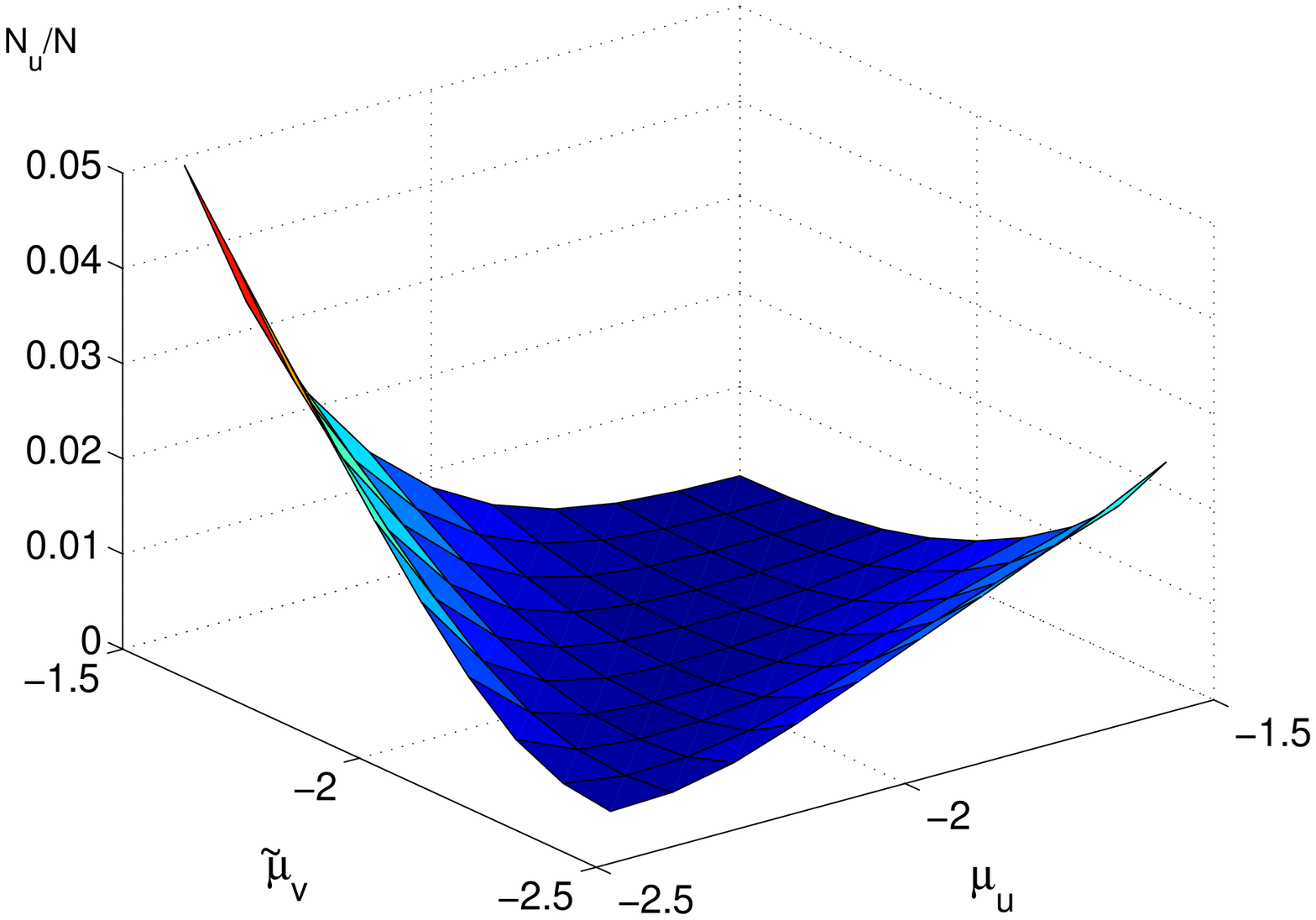}}%
\centering\subfigure[]{\includegraphics[width=3.0in]{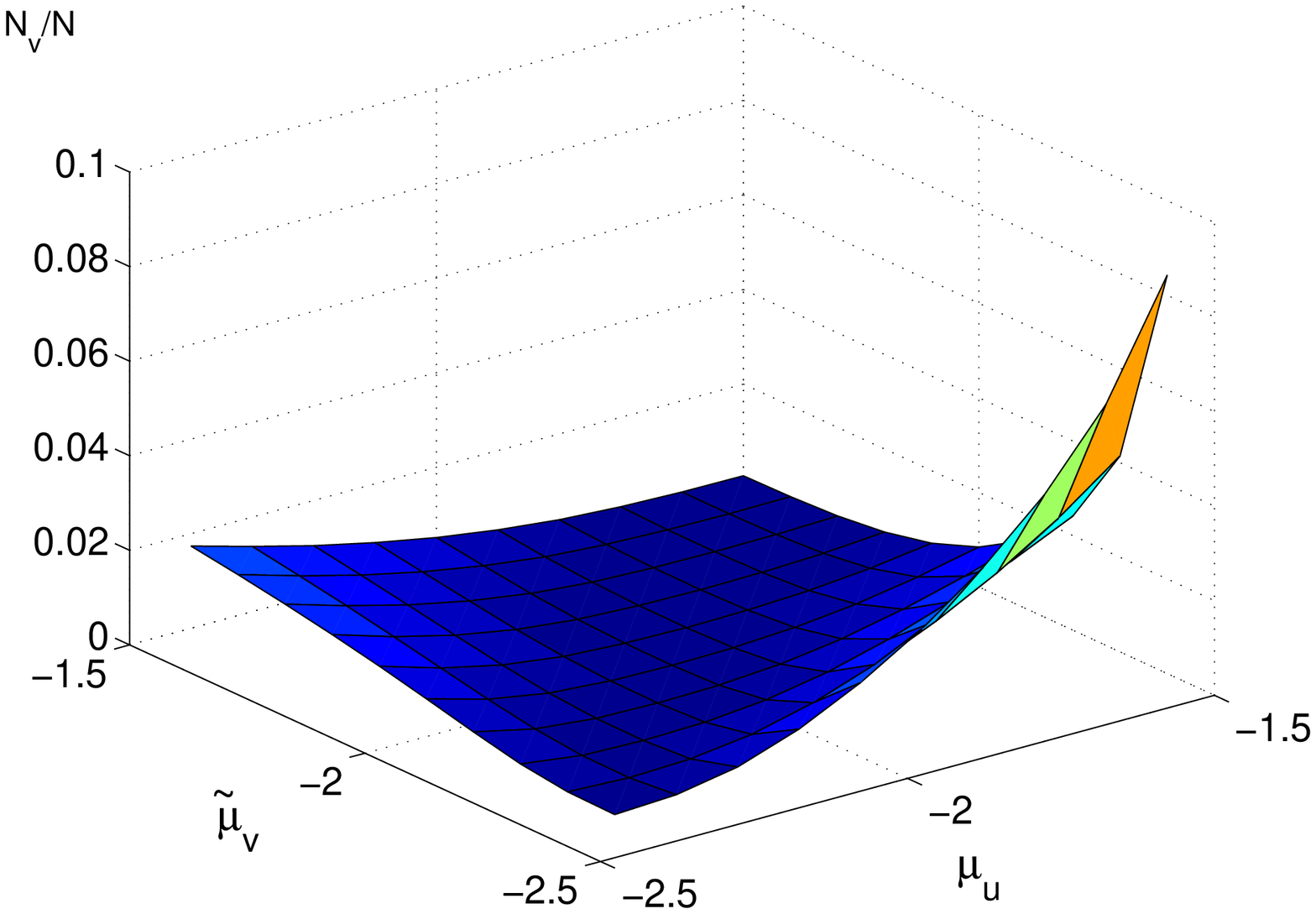}}
\par
\centering\subfigure[]{\includegraphics[width=3.0in]{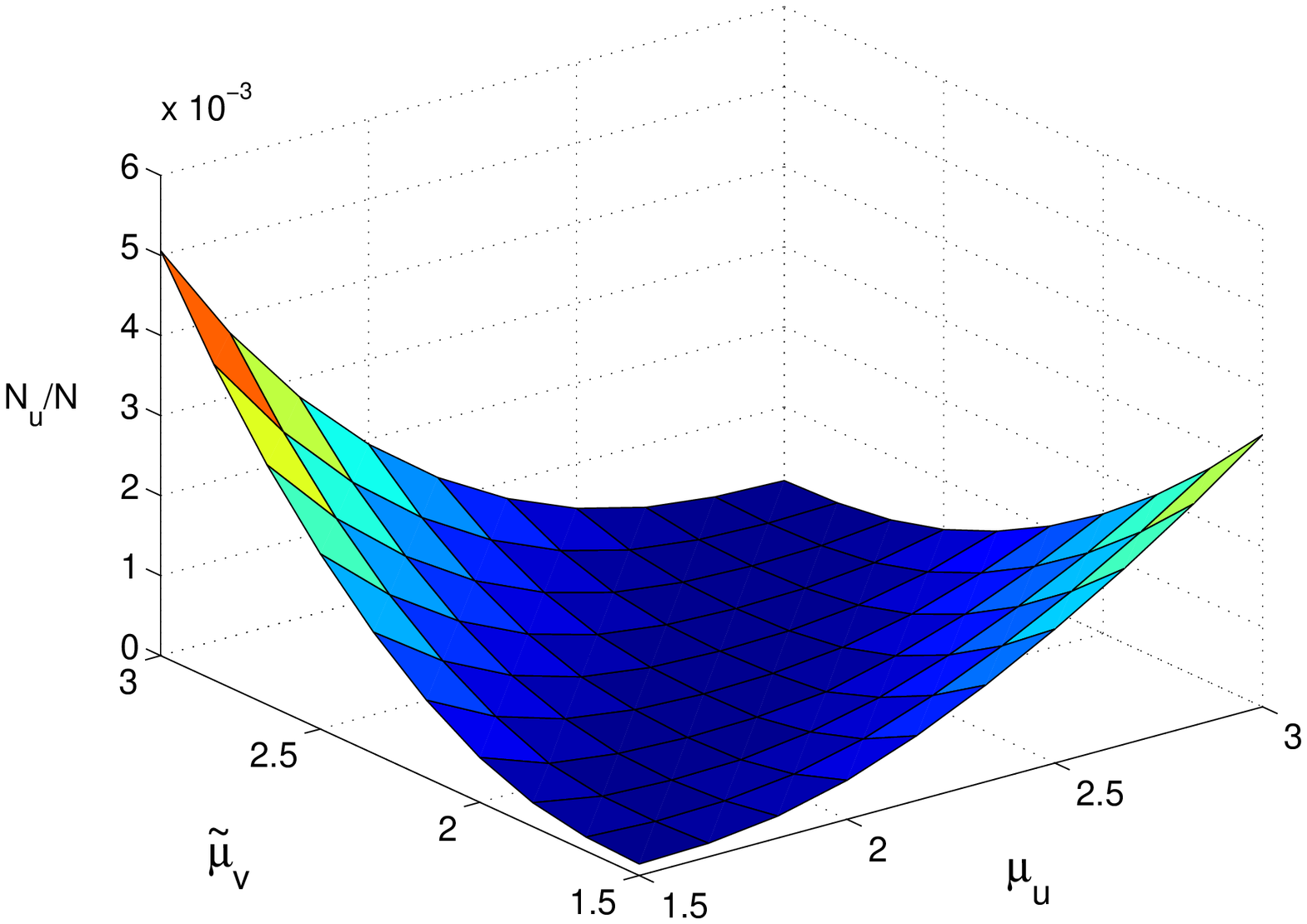}}%
\centering\subfigure[]{\includegraphics[width=3.0in]{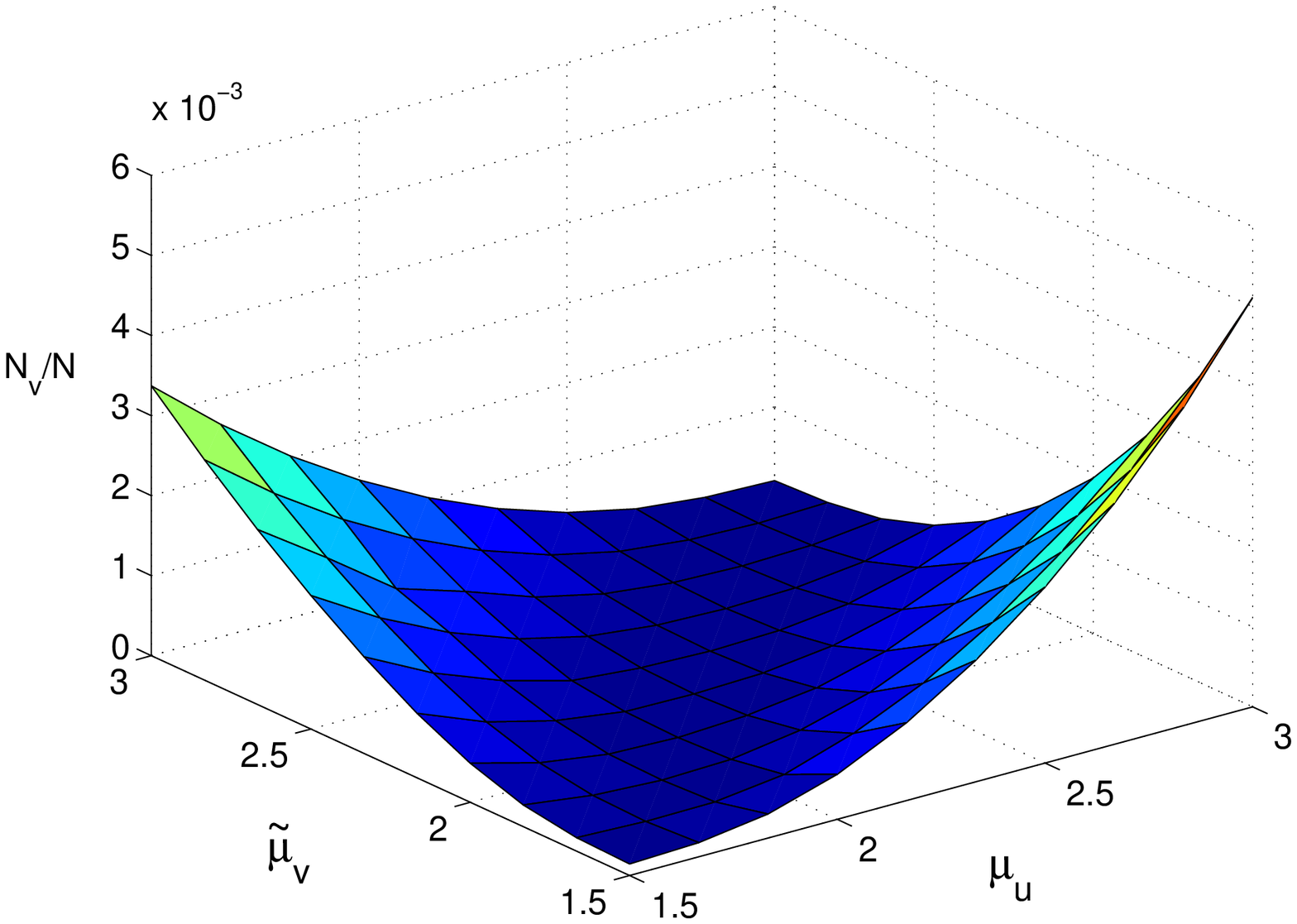}}
\caption{(Color online) The vapor fractions $N_{u}/N$ and $N_{v}/N$ in
families of stable tightly and loosely bound intragap solitons in the first
(a,b) and second (c,d) bandgaps versus $\protect\mu _{u}$ and $\tilde{%
\protect\mu}_{v}$, at $T=0$.}
\label{Fig2}
\end{figure}

Families of partially incoherent GSs can be characterized by fractions of
the vapor components in the total number of atoms, i.e., $N_{u}/N$ and $%
N_{v}/N$ [see Eq.~(\ref{N})], as functions of $\mu _{u}$ and $\tilde{\mu}_{v}
$, each chemical potential varying within a given bandgap. In Fig.~\ref{Fig2}
we display these dependences for $T=0$, with both $\mu _{u}$ and $\tilde{\mu}%
_{v}$ belonging to the first or second bandgap [then $\mu _{\phi }=\left(
\mu _{u}+\tilde{\mu}_{v}\right) /2$, see Eq. (\ref{locking}), lies in the
same gap, hence the families are of the \textit{intragap} type, indeed]. A
\textquotedblleft valley" in the plots running along the diagonal means that
the symmetric solutions, with $\mu _{u}=\tilde{\mu}_{v}$, amount to the
ordinary GSs, with $u=v=0$. All solitons belonging to these families in the
first and second gaps feature, respectively, tightly- and loosely-bound
shapes, similar to those in Fig.~\ref{Fig1}.

\begin{figure}[tbp]
\centering{\includegraphics[width=3.0in]{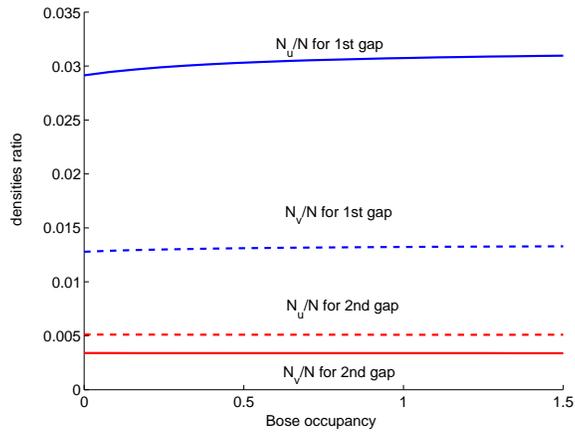}}
\caption{(Color online) The vapor fractions $N_{u}/N$ and $N_{v}/N$
vs.~temperature [shown via Bose occupancy $\mathcal{N}_{B}$, see Eq. (%
\protect\ref{Bose})] for the same solitons as in Fig.~\protect\ref{Fig1}.}
\label{Fig3}
\end{figure}

Figure~\ref{Fig3} shows that the dependence of $N_{u}/N$ and $N_{v}/N$ on
temperature is very weak, at fixed values of $\mu _{u}$ and $\tilde{\mu}_{v}$
(in the temperature range considered). Counterparts of the latter dependence
for fixed $N$ (rather than fixed chemical potentials) may be interesting
too, but they need a large pool of numerical data and will be reported
elsewhere.

\subsection{Intergap and higher-order solitons}

\begin{figure}[tbp]
\centering{\includegraphics[width=3.0in]{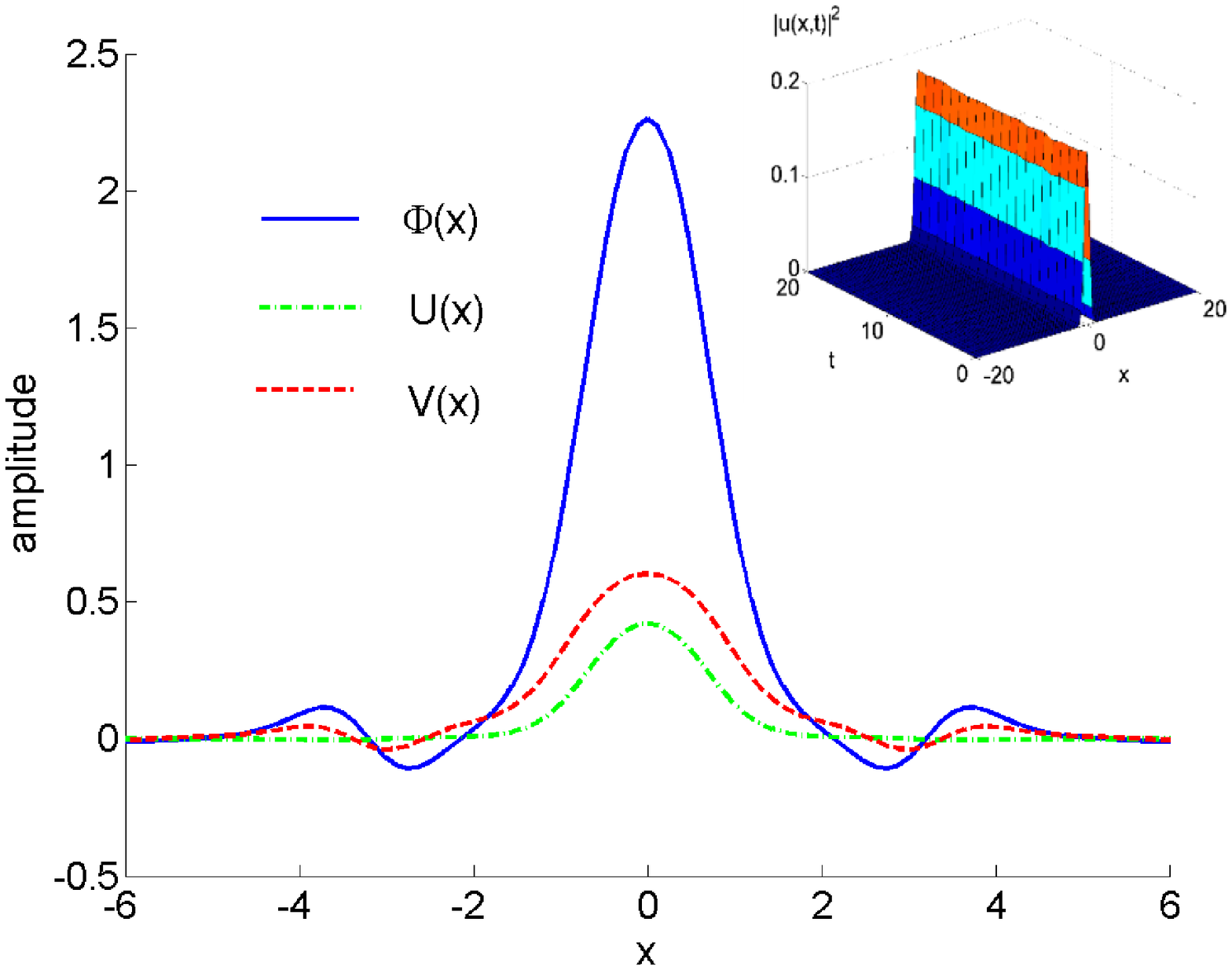}}
\caption{(Color online) An intergap soliton at $T=0$, with chemical
potentials $\protect\mu _{u}=0$ and $\tilde{\protect\mu}_{v}=3$, $\protect%
\mu _{\protect\phi }=1.5$, which fall in the first and second bandgaps,
respectively. The inset illustrates stability of the perturbed intergap
soliton in direct simulations.}
\label{Fig4}
\end{figure}

A family of \textit{intergap solitons} has been constructed too, with the
chemical potentials of the fluctuational components, $\mu _{u}$ and $\tilde{%
\mu}_{v}$, belonging to the first and second bandgaps, respectively [then,
the condensate's chemical potential, locked to $\mu _{u}$ and $\tilde{\mu}%
_{v}$ as per Eq. (\ref{locking}), $\mu _{\phi }=\left( \mu _{u}+\tilde{\mu}%
_{v}\right) /2$, belongs to the \emph{second} bandgap too]. A typical
soliton of this type is shown in Fig.~\ref{Fig4}. Note that its $U(x)$
component, which corresponds to the chemical potential, $\mu _{u}$, which
belongs to the first bandgap, features a typical tightly bound shape, cf.
Fig. \ref{Fig1}(a), while the shapes of the other two components, $\Phi (x)$
and $V(x)$, which pertain to the chemical potentials, $\mu _{\phi }$ and $%
\tilde{\mu}_{v}$, which belong to the second bandgap, are weakly bound, cf.
Fig. \ref{Fig1}(a).

\begin{figure}[tbp]
\centering\subfigure[]{\includegraphics[width=3.0in]{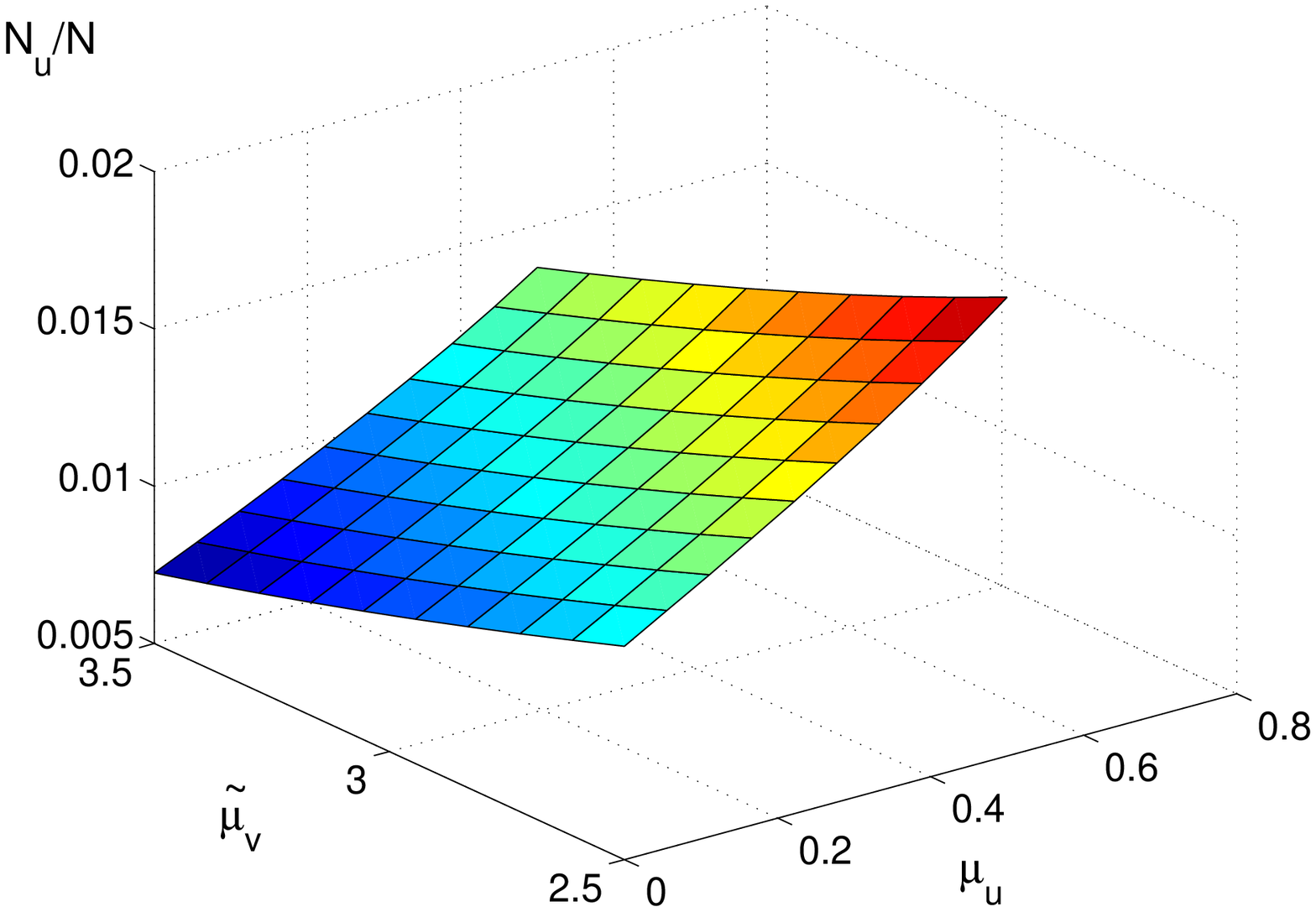}}%
\centering\subfigure[]{\includegraphics[width=3.0in]{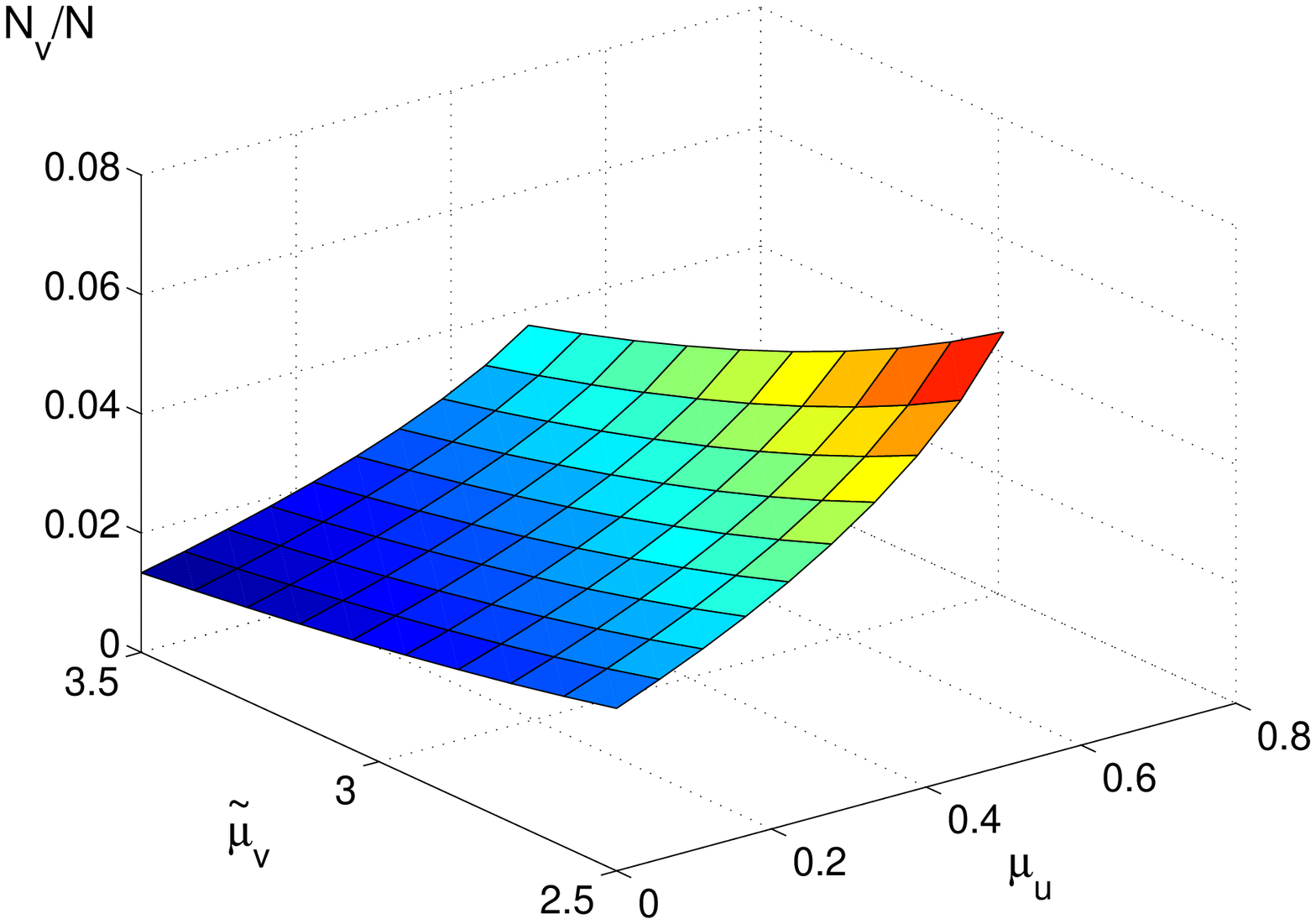}}
\par
\caption{(Color online) The vapor fractions $N_{u}/N$ and $N_{v}/N$ in the
family of intergap solitons, at $T=0$, versus chemical potentials of the two
fluctuational components, $\protect\mu _{u}$ and $\tilde{\protect\mu}_{v}$,
belonging to the first and second bandgaps, respectively.}
\label{Fig4extra}
\end{figure}

The $N_{u}/N$ and $N_{u}/N$ characteristics for the entire intergap family
are presented in Fig. \ref{Fig4extra}. Note that intergap solitons with $%
u=v=0$ do not exist, unlike their intragap counterparts, therefore these
plots do not feature \textquotedblleft valleys", unlike Fig. \ref{Fig2}.

\begin{figure}[tbp]
\centering
\includegraphics[width=3.0in]{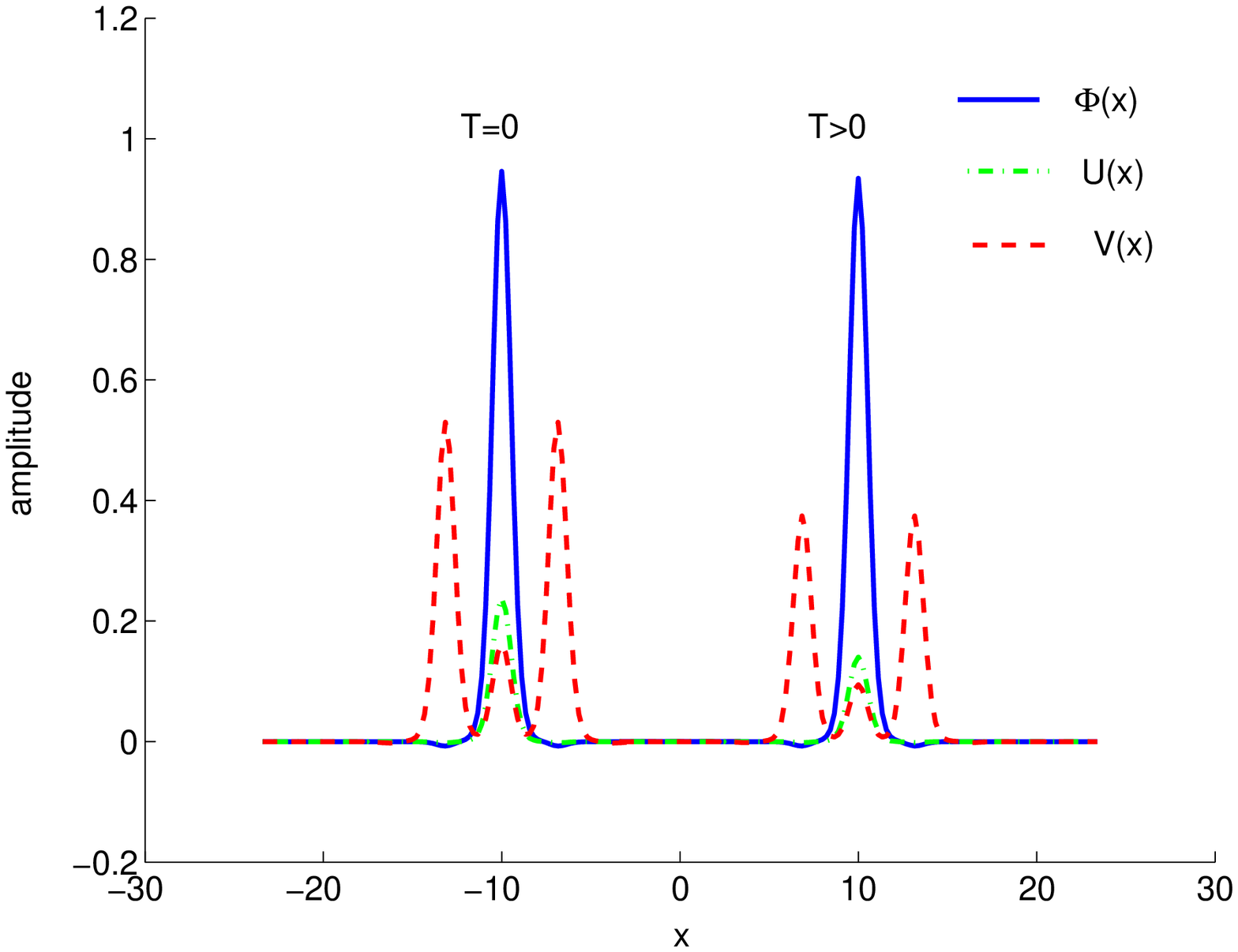}
\caption{Stable higher-order solitons in the first bandgap, for the same
parameters as in Fig. \protect\ref{Fig1}(a).}
\label{Fig5}
\end{figure}

In addition to the fundamental (single-humped) GSs, various types of
higher-order multi-humped states have been found too. In Fig.~\ref{Fig5}, we
display (for $T=0$ and $T>0$) the simplest among them, which is
single-humped in the condensate field ($\phi $), and doubled-humped in one
of the vapor components.

\begin{figure}[tbp]
\centering\subfigure[]{\includegraphics[width=3.0in]{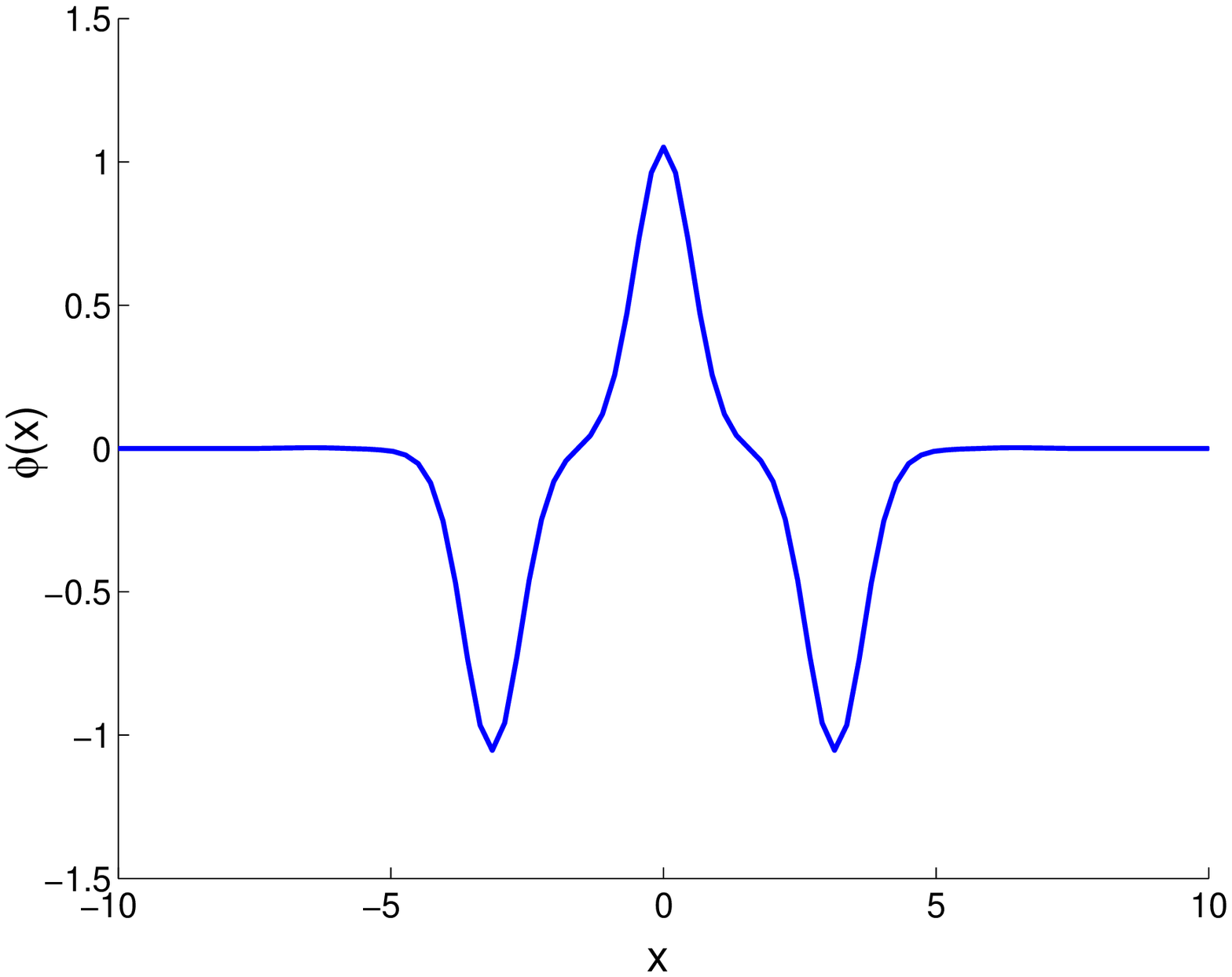}}%
\centering\subfigure[]{\includegraphics[width=3.0in]{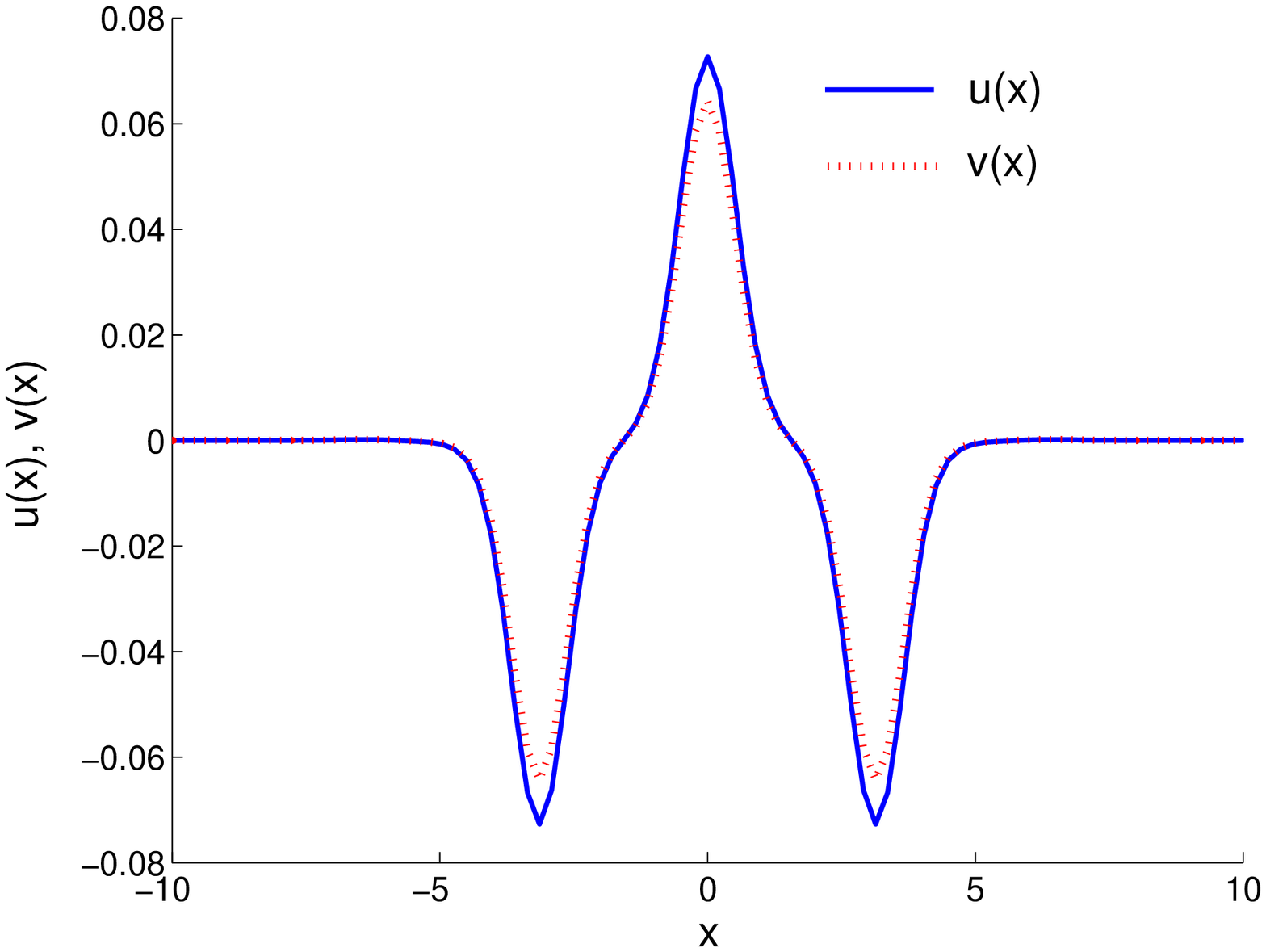}}
\par
\caption{(Color online) An example of a stable bound state of three
fundamental solitons belonging to the first bandgap (for $\protect\mu %
_{u}=-2.0$, $\tilde{\protect\mu}_{v}=-2.2$, $\protect\mu _{\protect\phi %
}=-2.1$): the condensate component (a) and vapor wave functions (b).}
\label{Fig5extra}
\end{figure}

Obviously, the higher-order soliton shown in Fig. \ref{Fig5} is not a bound
state of fundamental solitons. On the other hand, straightforward bound
states can be found too, see an example of a three-soliton complex in Fig. %
\ref{Fig5extra}. Note that, as per a general principle for the stability of
bound solitons on lattices \cite{Todd}, this complex may be stable because
the phase difference between the bound solitons is $\pi $.
As well as the fundamental intra- and intergap solitons, their higher-order
counterparts of various types form families which fill out the bandgaps.

\section{Stability analysis}

The stability of the GSs was first tested in direct simulations, which has
demonstrated that they are \emph{all} appear to be stable, both at $T=0$ and
$T>0$. Most accurate information about the stability can be obtained from
computation of eigenvalues for small perturbations, using equations (\ref%
{phi}) and (\ref{HFB_PDE}) linearized around stationary solitons. In
particular, the stability of ordinary GSs was previously shown in the
framework of the Bogoliubov--de Gennes equations, which are derived by the
linearization of the GP equation about the solitons \cite{Markus}.

Following this approach, we first consider the stability of the GSs with $%
u=v=0$ (i.e., the subfamily along the diagonal \textquotedblleft valleys" in
Fig. \ref{Fig2}) against small vapor perturbations, $u=e^{-i\lambda
t}u_{1}(x)$ and $v=e^{i\left( 2\mu _{\phi }-\lambda \right) t}v_{1}(x)$,
where $\lambda $ is the perturbation eigenvalue. The instability implies the
existence of eigenvalues with $\mathrm{Im}(\lambda )>0$. In this case,
linearized equations (\ref{HFB_PDE}) decouple from Eq.~(\ref{phi}), yielding
\begin{eqnarray}
(1/2)u_{1}^{\prime \prime }+\left[ \varepsilon \cos (2x)-2\rho \left\vert
\Phi \right\vert ^{2}\right] u_{1}+\Phi ^{2}v_{1} &=&\lambda u_{1},  \notag
\\
(1/2)v_{1}^{\prime \prime }+\left[ \varepsilon \cos (2x)-2\rho \left\vert
\Phi \right\vert ^{2}\right] v_{1}+\Phi ^{2}u_{1} &=&-\left( 2\mu _{\phi
}+\lambda \right) v_{1}.  \label{u0v0}
\end{eqnarray}%
Solving these equations (which do not depend on temperature) numerically
(with proper boundary conditions), we have concluded that all GSs with zero
vapor components are stable against \textquotedblleft vaporization".

Then, we performed the linear-stability analysis for the full GSs, including
nonzero vapor components. We have found that the families of intragap
solitons in both (first and second) bandgaps are completely stable (for $T=0$
and $T>0$ alike), in complete accordance with direct simulations.
Preliminary considerations of higher-order intragap solitons, such as ones
displayed in Figs. \ref{Fig5} and \ref{Fig5extra}, suggest that they are
stable too.

\begin{figure}[tbp]
\centering\includegraphics[width=3.0in]{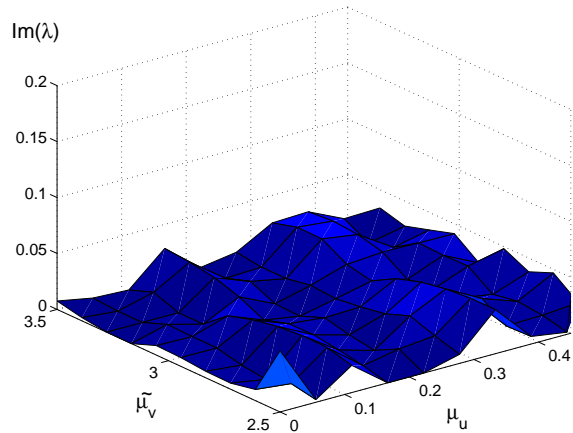}
\caption{The instability growth rate for the family of intergap solitons (at
$T=0$), as found from numerical solution of the linearized equations for
small perturbations.}
\label{Fig6}
\end{figure}

For the intergap family, a \emph{weak instability} is revealed by the
computation of eigenvalues, see Fig. \ref{Fig6} (since intergap solitons
cannot exist without vapor components, this instability is specific to the
partially incoherent GSs). However, this weak instability does not manifest
itself in direct simulations (as shown, for instance, by inset in Fig.~\ref%
{Fig4}), which suggests that the intergap solitons, even though being
formally unstable, may be observed in experiments.

It is also relevant to check the stability of the solitons against
deviations from the one-dimensionality \cite{Luca}. In the simplest
approximation, this amounts to modifying the equations, keeping them
effectively one-dimensional but adding quintic terms to the cubic
nonlinearity \cite{friction,Lev}. Preliminary analysis shows that no
additional instability emerges in this way.

\section{Conclusion}

We have found families of matter-wave gap solitons (GSs) in the degenerate
Bose gas with repulsive interactions between atoms, trapped at zero or
finite temperature in a periodic optical-lattice (OL) potential. Stability
of the GSs was studied too. The solitons include a coherent condensate wave
function, and two components of the incoherent \textquotedblleft vapor"
(which actually comprise many fluctuation modes, due to the OL's band
structure). Chemical potentials of all constituents of the GS must fall in
bandgaps. Accordingly, families of intra- and intergap solitons (including
higher-order ones, and bound states of fundamental solitons) were found in
the two lowest bandgaps, and it was concluded that they do not change
drastically with the growth of temperature. While the intragap GSs are
completely stable, their counterparts of the intergap type feature a very
small unstable perturbation eigenvalue, but, nevertheless, they do not
feature any tangible instability in direct simulations.

We acknowledge valuable discussions with A. Vardi. This work was supported
in part by Israel Science Foundation (Center-of-Excellence grant
No.~8006/03) and U.S.-Israel Binational Science Foundation (grant
No.~2002147).


\begin{thebibliography}{99}
\bibitem{HFB} A. Griffin, Phys. Rev. B \textbf{53}, 9341 (1996); N. P.
Proukakis and K. Burnett, J. Res. Natl. Inst. Stand. Technol. \textbf{101},
457 (1996); N. P. Proukakis, K. Burnett, and H. T. C. Stoof, Phys. Rev. A%
\textbf{57}, 1230 (1998); P. O. Fedichev and G. V. Shlyapnikov, Phys. Rev. A%
\textbf{58}, 3146 (1998); M. Holland, J. Park, and R. Walser, Phys. Rev.
Lett. \textbf{86}, 1915 (2001); A. M. Rey, B. L. Hu, E. Calzetta, A. Roura,
and C. W. Clark, Phys. Rev. A\textbf{69}, 033610 (2004).

\bibitem{Vardi} H. Buljan, M. Segev and A. Vardi, Phys. Rev. Lett. \textbf{95%
}, 180401 (2005).

\bibitem{depletion} M. Lewenstein and L. You, Phys. Rev. Lett. \textbf{77},
3489 (1996); Y. Castin and R. Dum, \textit{ibid}. \textbf{79}, 3553 (1997).

\bibitem{depletion-dark} J. Dziarmaga, Z. P. Karkuszewski, and K. Sacha,
Phys. Rev. A \textbf{66}, 043615 (2002); J. Dziarmaga and K. Sacha, \textit{%
ibid}. \textbf{66}, 043620 (2002); C. K. Law, \textit{ibid}. \textbf{68},
015602 (2003); A. E. Muryshev, G. V. Shlyapnikov, W. Ertmer, K. Sengstock,
and M. Lewenstein, Phys. Rev. Lett. \textbf{89}, 110401 (2002).

\bibitem{Canberra} R.-K. Lee , E. A. Ostrovskaya, Y. S. Kivshar, and Y. Lai,
Phys. Rev. A \textbf{72}, 033607 (2005).

\bibitem{Luca} L. Salasnich, J. Phys. B -- At. Mol. Opt. Phys.\textbf{39},
1743 (2006).

\bibitem{friction} S. Sinha , A. Yu. Cherny, D. Kovrizhin, and J. Brand ,
Phys. Rev. Lett. \textbf{96}, 030406 (2006).

\bibitem{exper-bright} L. Khaykovich, F. Schreck, G. Ferrari, T. Bourdel, J.
Cubizolles, L. D. Carr, Y. Castin, and C. Salomon, Science \textbf{296},
1290 (2002); K. E. Strecker, G. B. Partridge, A. G. Truscott, and R. G.
Hulet, Nature \textbf{417}, 150 (2002).

\bibitem{Cornish} S. L. Cornish, S. T. Thompson, and C. E. Wieman, Phys.
Rev. Lett. \textbf{96}, 170401 (2006).

\bibitem{Moti} M. Mitchell, Z. Chen, M. Shih, and M. Segev , Phys. Rev.
Lett. \textbf{77}, 490 (1996); M. Mitchell , M. Segev, T. H. Coskun and D.
N. Christodoulides , Phys. Rev. Lett. \textbf{79}, 4990 (1997); M. Mitchell
and M. Segev, Nature \textbf{387}, 880 (1997); D. N. Christodoulides , E. D.
Eugenieva, T. H. Coskun, M. Segev, M. Mitchell , Phys. Rev. E \textbf{63},
035601(R) (2001).

\bibitem{GSprediction} A. Trombettoni and A. Smerzi, Phys. Rev. Lett.
\textbf{86}, 2353 (2001); F. Kh. Abdullaev , B. B. Baizakov, S. A.
Darmanyan, V. V. Konotop and M. Salerno, , Phys. Rev. A \textbf{64,} 043606
(2001); I. Carusotto, D. Embriaco, and G. C. La Rocca, \textit{ibid}. 65,
053611 (2002); B. B. Baizakov, V. V. Konotop, and M. Salerno, J. Phys. B%
\textbf{35}, 5105 (2002); E. A. Ostrovskaya and Y. S. Kivshar, Phys. Rev.
Lett. \textbf{90}, 160407 (2003).

\bibitem{gap_sol} B. Eiermann \textit{et al}., Phys. Rev. Lett. \textbf{92},
230401 (2004).

\bibitem{Ahufinger} V. Ahufinger and A. Sanpera, Phys. Rev. Lett. 94, 130403
(2005).

\bibitem{Gubeskys} A. Gubeskys, B. A. Malomed, and I. M. Merhasin, Phys.
Rev. A\textbf{73}, 023607 (2006).

\bibitem{Moti2} O. Cohen, T. Schwartz, J. W. Fleischer, M. Segev, and D. N.
Christodoulides, Phys. Rev. Lett. 91, 113901 (2003).

\bibitem{Sukhorukov} A. A. Sukhorukov and Y. S. Kivshar, Phys. Rev. Lett.
\textbf{91}, 113902 (2003).

\bibitem{3Dto1D} V. M. P\'{e}rez-Garc\'{\i}a, H. Michinel, and H. Herrero,
Phys. Rev. A \textbf{57}, 3837 (1998); L. Salasnich, A. Parola, and L.
Reatto, Phys. Rev. A \textbf{65}, 043614 (2002); A. E. Muryshev, G. V.
Shlyapnikov, W. Ertmer, K. Sengstock, and M. Lewenstein, Phys. Rev. Lett.
\textbf{89}, 110401 (2002); Y. B. Band, I. Towers, and B. A. Malomed, Phys.
Rev. A \textbf{67}, 023602 (2003); S. Sinha, A. Y. Cherny, D. Kovrizhin, and
J. Brand, Phys. Rev. Lett. \textbf{96}, 030406 (2006).

\bibitem{Lev} L. Khaykovich and B. A. Malomed, Phys. Rev. A, in press
(article No. AS10005).

\bibitem{Todd} T. Kapitula, P. G. Kevrekidis, and B. A. Malomed,. Phys. Rev.
E \textbf{63}, 036604.

\bibitem{Markus} K. M. Hilligs\o e, M. K. Oberthaler, K. P. Marzlin, Phys.
Rev. A\textbf{66}, 063605 (2002); D. E. Pelinovsky, A. A. Sukhorukov, and Y.
S. Kivshar, Phys. Rev. E \textbf{70}, 036618 (2004).
\end{thebibliography}
\end{document}